# HERITABILITY ESTIMATES ON RESTING STATE FMRI DATA USING THE ENIGMA ANALYSIS PIPELINE

Bhim M. Adhikari
Maryland Psychiatry Research Center, Department of Psychiatry, University of Maryland School of Medicine, Baltimore, MD, USA
Email: badhikari@mprc.umaryland.edu

Neda Jahanshad
Imaging Genetics Center, Stevens Neuroimaging and Informatics Institute,
Keck School of Medicine of USC, Marina del Rey, CA, USA
Email: neda.jahanshad@usc.edu

Dinesh Shukla
Maryland Psychiatry Research Center, Department of Psychiatry, University of Maryland School of Medicine, Baltimore, MD, USA
Email: dshukla@mprc.umaryland.edu

David C. Glahn
Department of Psychiatry, Yale University, School of Medicine, New Haven, CT, USA
Email: david.glahn@yale.edu

John Blangero
Genomics Computing Center, University of Texas at Rio Grande Valley
Email: john.blangero@UTRGV.edu

Richard C. Reynolds
National Institute of Mental Health, Bethesda, MD, USA
Email: reynoldr@mail.nih.gov

Robert W. Cox
National Institute of Mental Health, Bethesda, MD, USA
Email: robertcox@mail.nih.gov

Els Fieremans
Center for Biomedical Imaging, Department of Radiology, New York University School of Medicine, NY, USA
Email: els.fieremans@nyumc.org

Jelle Veraart
Center for Biomedical Imaging, Department of Radiology, New York University School of Medicine, NY, USA
Email: Jelle.Veraart@nyumc.org

Dmitry S. Novikov
Center for Biomedical Imaging, Department of Radiology, New York University School of Medicine, NY, USA
Email: Dmitry.Novikov@nyumc.org

L. Elliot Hong
Maryland Psychiatry Research Center, Department of Psychiatry, University of Maryland School of Medicine, Baltimore, MD, USA
Email: ehong@mprc.umaryland.edu

Paul M. Thompson
Imaging Genetics Center, Stevens Neuroimaging & Informatics Institute, Keck School of Medicine of USC, Marina del Rey, CA, USA
Email: pthomp@usc.edu

Peter Kochunov
Maryland Psychiatry Research Center, Department of Psychiatry, University of Maryland School of Medicine, Baltimore, MD, USA
Email: pkochunov@mprc.umaryland.edu

Big data initiatives such as the Enhancing NeuroImaging Genetics through Meta-Analysis consortium (ENIGMA), combine data collected by independent studies worldwide to achieve more generalizable estimates of effect sizes and more reliable and reproducible outcomes. Such efforts require harmonized image analyses protocols to extract phenotypes consistently. This harmonization is particularly challenging for resting state fMRI due to the wide variability of acquisition protocols and scanner platforms; this leads to site-to-site variance in quality, resolution and temporal signal-to-noise ratio (tSNR). An effective harmonization should provide optimal measures for data of different qualities. We developed a multi-site rsfMRI analysis pipeline to allow research groups around the world to process rsfMRI scans in a harmonized way, to extract consistent and quantitative measurements of connectivity and to perform coordinated statistical tests. We used the single-modality ENIGMA rsfMRI preprocessing pipeline based on model-free Marchenko-Pastur PCA based denoising to verify and replicate resting state network heritability estimates. We analyzed two independent cohorts, GOBS (Genetics of Brain Structure) and HCP (the Human Connectome Project), which collected data using conventional and connectomics oriented fMRI protocols, respectively. We used seed-based connectivity and dual-regression approaches to show that the rsfMRI signal is consistently heritable across twenty major functional network measures. Heritability values of 20-40% were observed across both cohorts.

## 1. Introduction

Resting state functional MRI (rsfMRI) studies investigate large-amplitude, spontaneous low-frequency fluctuations in the fMRI signal that are temporally correlated across functionally related brain areas [1-4]. It is the basis for a powerful method to evaluate temporal correlations of low-frequency blood oxygenation level-dependent fluctuations across brain regions in the absence of a task or stimulus [2, 4]. Genetic analyses on rsfMRI phenotypes are challenged by limited statistical power. One way to address this is by pooling data from multiple cohorts or studies. The Enhancing Neuroimaging Genetics through Meta-Analysis (ENIGMA) consortium has developed an rsfMRI analysis pipeline to perform consistent analysis and extraction of resting state connectivity measures across data collected using diverse protocols [7]. Here, we demonstrate the utility of this pipeline by replicating findings of significant heritability in the default mode network in (A) the original Genetics of Brain Structure (GOBS) cohort [10], and (B) data from the Human Connectome Project (HCP; [11]); we also aim to demonstrate consistent additive genetic contribution to intersubject variance in other intrinsic brain networks. The motivation for this work is that, ultimately, it should be possible to discover genetic variants that reliably affect brain function, but a key milestone in this quest is to establish that the metrics targeted are heritable, i.e., individual genetic variance accounts for a significant proportion of their variation across subjects. This paper is one key step in this process.

The ENIGMA rsfMRI pipeline differs from existing pipelines in two notable respects. Many fMRI analysis pipelines require input from multiple imaging modalities. Commonly, a structural T1-weighted (T1w) MRI scan is required to regress out signal trends from cerebrospinal fluid (CSF) and cerebral white matter, and T1w data is used for anatomical registration to an atlas space [5, 6]. In the spirit of other ENIGMA pipelines, our pipeline for rsfMRI is a single-modality pipeline. It uses a deformable template created from 1,100 individual images provided by ENIGMA sites to incorporate shape distortions common to fMRI images [7]. The pipeline uses direct tissue classification of rsfMRI data to regress out some of the variance due to methodical (non-biological) factors. Both approaches avoid the potential pitfalls of site-to-site variance in T1w data and coregistration biases that may influence the rsfMRI phenotypes.

In another notable difference, we use a novel denoising technique based on the Marchenko-Pastur distribution [8] - fitted to the eigenvalues of a principal component analysis (MP-PCA) across space and time - to identify and remove the principal components originating due to thermal noise. We used this MP-PCA-based denoising technique to reduce signal fluctuations rooted in thermal noise and hence increase the tSNR without altering the spatial resolution. The ability to suppress thermal noise is based on data redundancy in the PCA domain, using universal properties of the eigenspectrum of random covariance matrices [8]. The bulk of the PCA eigenvalues arise due to noise and can be asymptotically represented by the universal MP distribution, in the limit of the large signal matrix size (voxels × time points). The Marchenko-Pastur parameterization allows us to identify noise-only components and to estimate the noise level in a local neighborhood based on the singular value decomposition of a signal matrix combining neighborhood voxels [9]. After removing noise-only components, the resulting images show enhanced SNR; the residual noise is contained in the remaining components, which cannot be further denoised by this method.

Here, we validate the ENIGMA-rsfMRI analysis pipeline by attempting to verify and replicate the assertion that there is moderately strong genetic influence on the resting state signal [10], and that significant heritability estimates are consistently found across independent cohorts.

While the actual value of the heritability estimate may depend on cohort parameters such as their demographics, age, and environment – and presumably also on the image SNR – a key goal of a collaborative imaging genetics initiative is to identify brain metrics that are significantly heritable across cohorts, as a precursor to a more in-depth search for common variants associated with the trait, in this case brain function. Glahn and colleagues showed that individual variance in measures of connectivity within intrinsic brain networks - such as the default-mode network - is influenced by genetic factors: for some metrics, ~40% of the variance could be attributed to additive genetic factors [10]. The ENIGMA-rsfMRI pipeline was used to measure individual variations in the default mode network and perform heritability analysis in the same dataset used by Glahn and colleagues (N=334), using two complementary measurements of connectivity: region-based (also called "seed"-based) analysis and dual regression. We further expanded this analysis by demonstrating that similar heritability estimates can be obtained for other intrinsic brain networks and replicated the detection of significant heritability estimates in the data from a young adult sample collected by HCP (N=518). The detection and estimation of additive genetic effects may depend on the degree of relatedness across individuals underlying the sample structure. We tested heritability measurements computed from two commonly used familial study designs: GOBS subjects were recruited from an extended pedigree and HCP subjects were recruited from a twin/siblings registry. Similarity in the heritability measurements across two diverse cohorts would therefore be further evidence to support the suitability of the ENIGMA rsfMRI protocol and connectivity measurements for large-scale genetic analyses of cerebral functional connectivity.

## 2. Methods and Materials

### 2.1. Study subjects and imaging protocols:

Two rsfMRI datasets were analyzed (GOBS and HCP: acronyms are detailed below).

#### 2.1.1. GOBS - Genetics of Brain Structure and Function study

Subjects: This sample comprised 334 (124 M/210 F, mean age: 47.9±13.2 years) Mexican-American individuals from 29 extended pedigrees (average family size = 9 people; range 5-32) who participated in the Genetics of Brain Structure and Function study. Individuals in this cohort have actively participated in genetics research for over 20 years and, were randomly selected from the community with the constraints that they are of Mexican-American ancestry, part of a large family, and live within the San Antonio, TX region. In this study, individuals were excluded for MRI contraindications, history of neurological illnesses, or stroke or other major neurological event. All participants provided written informed consent on forms approved by the institutional review board at the University of Texas Health Science Center San Antonio (UTHSCSA).

Imaging: All imaging was performed at the Research Imaging Institute, UTHSCSA, on a Siemens 3 T Trio scanner using a multichannel phased array head coil. Whole-brain, resting-state functional imaging was performed using a gradient-echo echo planer imaging (EPI) sequence sensitive to the BOLD effect with the following parameters: TR=3000 ms, TE=30 ms, spatial resolution=1.72×1.72×3 mm$^3$, flip angle=90 degrees. The resting-state protocol included 43 slices acquired parallel to the sagittal plane containing the anterior and posterior commissures; scan time was 7.5 min.

*2.1.2. HCP – Human Connectome Project*

Subjects: We included rsfMRI data, 518 participants (240 M/278 F; mean age 28.7 ± 3.7 years) from the Human Connectome Project (HCP) dataset, released in March 2017. Participants were recruited from the Missouri Family and Twin Registry [11]. All HCP participants were from young adult sibships of average size 3–4 that include a monozygotic or dizygotic twin pair and (where available) their non-twin siblings. Subjects ranged in age from 22 to 37 years. This age range was chosen as it corresponds to a period after neurodevelopment is largely completed and before the typical age of onset of neurodegenerative changes. The inclusion and exclusion criteria are detailed elsewhere [11]. The HCP subjects are healthy young adults within a restricted age range and free from major psychiatric or neurological illnesses [12, 13]. All subjects provided written informed consent on forms approved by the Institutional Review Board of Washington University in St Louis.

Imaging: All HCP subjects are scanned on a customized Siemens 3 T "Connectome Skyra" scanner housed at Washington University in St. Louis, using a standard 32-channel Siemens receive head coil. RsfMRI data consisted of two runs in one session. Within a session, oblique axial acquisitions alternated between phase encoding in a right-to-left direction in one run and phase encoding in a left-to-right direction in the other run. Resting state images were collected using a gradient-echo echo planar imaging (EPI) sequence with the following parameters: TR=720 ms, TE=33.1 ms, flip angle=52 degrees, FOV=208×180 mm (RO×PE), matrix=104×90 (RO×PE), 2.0 mm isotropic voxels, 72 axial slices, multiband factor=8; scan time was 28.8 min.

*2.2. Functional Image Analysis*

The rsfMRI data processing was carried out using the ENIGMA resting state analysis pipeline implemented in the Analysis of Functional NeuroImages (AFNI) software [14]. ENIGMA developed a single-modality resting state analysis pipeline [7]. The ENIGMA pipeline is an extension of the conventional AFNI rsfMRI pipeline [14] (**Figure 1**). The first step is the application of principal components analysis (PCA)-based denoising [8, 9], to improve signal-to noise ratio (SNR) and temporal SNR (tSNR) properties of the time series data, with no loss of spatial resolution of the image and without the introduction of additional partial volume effects [7]. This denoising approach is free from the limitations of the loss of spatial resolution of the image and introduction of additional partial volume effects that lead to complications in further quantitative analyses [8]. The MP-PCA approach does not alter the resting state network activation patterns, whereas spatial smoothing using a Gaussian kernel leads to partial voxel averaging, spreading the activations across gray and white matter regions and removing smaller nodes. Finally, the noise-maps produced by MP-PCA approach provide valuable information for quality control as deviations from the expected uniform or slowly varying in space pattern of thermal noise may indicate problems with the coil or other scanner hardware.

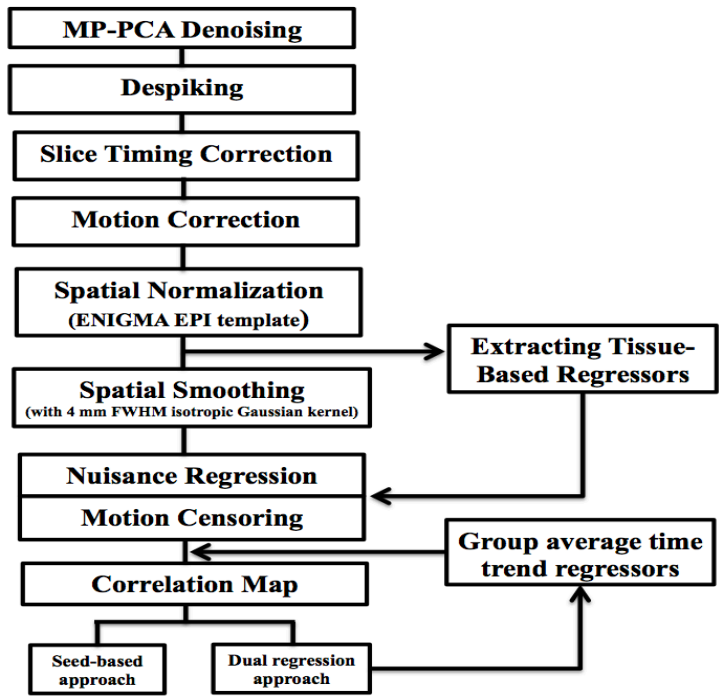

Fig. 1. Flowchart of ENIGMA rsfMRI analysis pipeline.

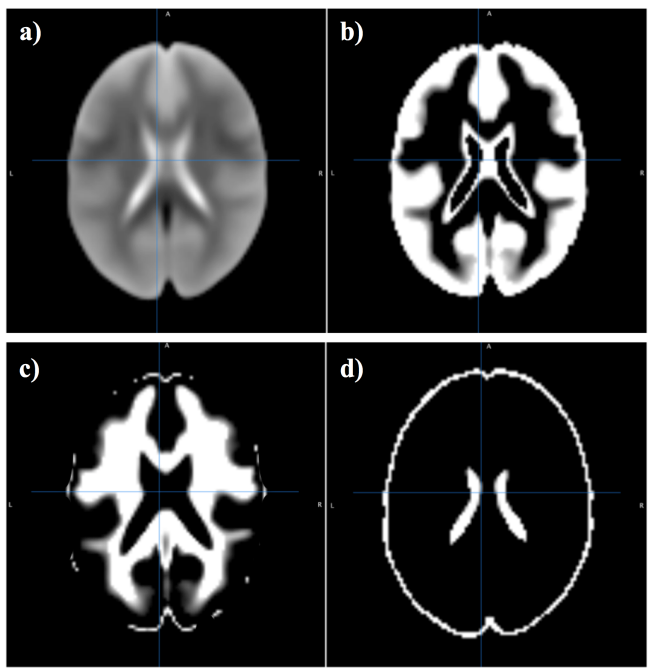

Fig. 2. ENIGMA EPI brain template (a) and segmented tissue classes (b-d) for gray matter, white matter and cerebrospinal fluid, respectively.

In the next step, supplementary data, if provided, is used for correction of spatial distortions associated with long-TE gradient echo imaging. Two available corrections are the gradient-echo 'fieldmap' or the reversed-gradient approach. In the next step, the data is registered to the ENIGMA EPI template (**Figure 2**) that was derived from 1,100 datasets corrected across 22 sites [7] to develop a spatial template and spatial atlas. This atlas has a dual purpose: it is used for regression of the global signal, and also offers a common anatomical spatial reference frame. Next, correction for head motion is performed by registering each functional volume to the first time point of the run. Nuisance variables such as the linear trend, 6 motion parameters (3 rotational and 3 translational directions), their 6 temporal derivatives (rate of change in rotational and translational motion) and time courses from the white matter and cerebrospinal fluid (CSF) from lateral ventricles were modeled using multiple linear regression analysis, which were then removed as regressors of no interest. Time points with excessive motion (> 0.2 mm), estimated as the magnitude of displacement from one time point to the next, including neighboring time points and outlier voxels fraction (>0.1) were censored from statistical analysis. Images were spatially normalized to the ENIGMA EPI template in Montreal Neurological Institute (MNI) standard space for group analysis.

## 2.3. Functional connectivity analysis

Resting state network templates were defined based on the probabilistic regions of interest (ROIs) from 20-component analysis of the BrainMap activation database and resting fMRI dataset [4]. We defined the binary masks of the resting state template regions from auditory network (AN), default mode network (DMN), fronto-parietal network (FPN), sensorimotor network (SMN), visual network (VN), executive control network (ECN), salience network (SN), and attention network (AttN) (**Figure 3**).

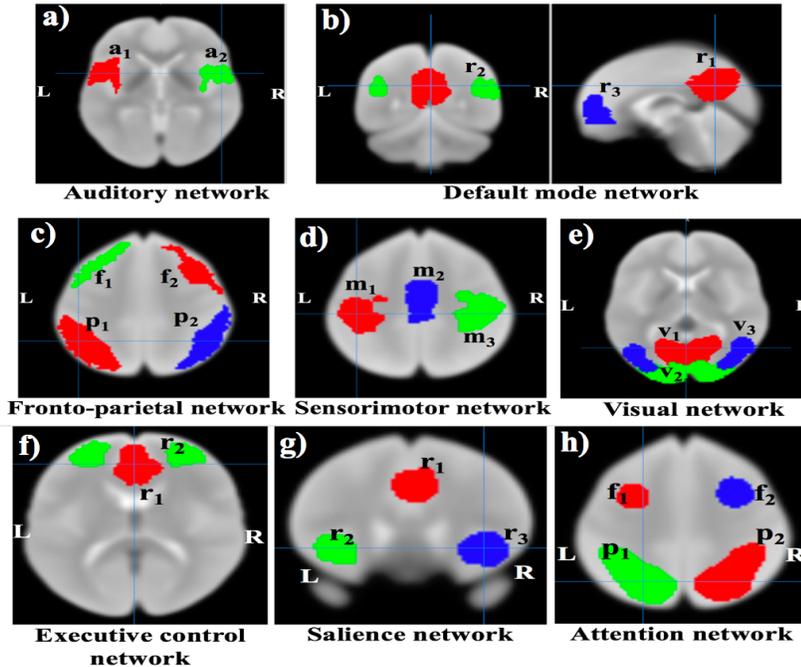

Fig.3. Resting state network template ROIs based on the BrainMap activation database and resting fMRI dataset [4]. Here, L=left, R=right, in (a) a1=left primary and association auditory cortices, a2=right primary and associated auditory cortices, in (b) r1=posterior cingulate/precuneus, r2=bilateral temporal-parietal regions and, r3=ventromedial frontal cortex, in (c) f1/f2=left/right frontal area and p1/p2=left/right parietal area, in (d) m1/m3=left/right motor area and m2=supplementary motor area, in (e) v1=medial visual areas, v2= occipital visual areas, and v3=lateral visual areas, in (f) r1=anterior cingulate cortex and r2=bilateral medial frontal gyrus, in (g) r1=anterior cingulate cortex and r2/r3=left/right insula, in (h) f1/f2=left/right middle frontal gyrus and p1/p2=left/right superior parietal lobule.

Mean time series were extracted from the seed regions of each network and connectivity maps corresponding to each seed region were obtained by assessing correlations along the time series for different regions. Next, Fisher's *r*-to-*z* transformations were applied to obtain a normal distribution. We calculated seed-based functional connectivity values between seed regions in each network and performed heritability calculation. For the HCP dataset, heritability measures were calculated for all subjects (N=518) under consideration and, in a separate analysis, for subjects (N=481) with censored TRs less than 15% of the total TRs during the processing. Furthermore, we performed dual regression analysis for the default mode network template ROIs, and calculated the functional connectivity measures and hence heritability estimates on the GOBS dataset (N=334). (This same dataset was used in a prior study by David Glahn and colleagues [10]). In the case of dual regression, for the given network template ROIs, average single time series were computed from the preprocessed data for each subject, and hence we obtained the average time series from all subjects, and then averaged these to obtain an average time series that represents the group average time trend. The group average time trend was regressed out from each subject's data, before calculating the functional connectivity values.

## 2.4. Heritability estimation

For the heritability estimations, the variance components method was used, as implemented in the Sequential Oligogenic Linkage Analysis Routines (SOLAR) Eclipse software package (http://www.nitrc.org/projects/se_linux) [15]. SOLAR uses maximum likelihood variance decomposition methods, extensions of the strategy developed by Amos and colleagues [16]. The covariance matrix Ω for a pedigree is given by: $\Omega = 2\Phi\sigma_g^2 + I\sigma_e^2$, where $\sigma_g^2$ is the genetic variance due to the additive genetic factors, $\Phi$ is the kinship matrix representing the pair-wise kinship coefficients among all individuals, $\sigma_e^2$ is the variance due to individual — unique environmental effects and measurement error, and I is an identity matrix (under the assumption that all environmental effects are uncorrelated among family members). Narrow sense heritability is defined as the fraction of phenotypic variance $\sigma_p^2$ attributable to additive genetic factors,

$$h^2 = \frac{\sigma_g^2}{\sigma_p^2} \qquad (1)$$

The variance parameters are estimated by comparing the observed phenotypic covariance matrix with the covariance matrix predicted by kinship [15]. Significance of the heritability estimate is tested by comparing the likelihood of the model in which $\sigma_g^2$ is constrained to zero with that of a model in which $\sigma_g^2$ is estimated. Twice the difference between the log-likelihoods of these models yields a test statistic, which is asymptotically distributed as a 1/2:1/2 mixture of $\chi^2$ variables with 1 degree-of-freedom and a point mass at zero. Prior to the heritability estimation, phenotype values from each dataset were adjusted for covariates including sex, age, age$^2$, age×sex interaction, and age$^2$×sex interaction. Inverse Gaussian transformation was also applied to ensure normality of the distribution. Outputs from SOLAR include the heritability estimate (h$^2$), the significance value (p), and the standard error for each trait (SE).

## 3. Results

*3.1. Seed-based analysis.* Heritability estimates for connectivity measurements extracted from the seed-based approach are summarized in **Table 1**. The default mode network (DMN) showed significant heritability for the connectivity from the posterior cingulate/precuneus to bilateral temporal-parietal regions ($h^2=0.34\pm0.16$, p=0.014) and ventromedial frontal cortex ($h^2=0.35\pm0.17$, p=0.014) respectively for the GOBS dataset. Replication analyses, in the HCP dataset, demonstrated significant heritability in all nodes of the DMN. In the fronto-parietal network, we found significantly heritable functional connectivity in both datasets. Heritability estimates in other networks showed a similar pattern of genetic control in both GOBS and HCP with greater evidence for statistical significance (i.e., lower p-values) observed in HCP subjects. Heritability estimates were found to be improved by excluding subjects who had more than 15% of total TRs censored due to motion from the HCP dataset (HCP, N=481).

*3.2. Dual regression analysis.* To confirm the heritability estimates from seed based connectivity, a dual regression analysis was also conducted in GOBS. Here, connectivity values for the DMN ROIs were again significantly heritable from posterior cingulate/precuneus to bilateral temporal-parietal regions ($h^2=0.31\pm0.17$, p=0.027) and ventromedial frontal cortex ($h^2=0.25\pm0.17$, p=0.038) **(Table 2)**. The connection from bilateral temporal-parietal regions to posterior cingulate/precuneus was likewise significantly heritable ($h^2=0.26\pm0.16$, p=0.035).

Table 1. Heritability estimates for measures derived from resting state networks (RSNs). *Regions are based off of **Figure 3**. Bolded connections are significant at 5% FDR. #Estimated heritability, $h^2$ (SE). Abbreviations: GOBS= Genetic of Brain Structure and Function study, HCP=Human Connectome Project, DMN=default mode network, FPN=fronto-parietal network, SMN=sensorimotor network, VN=visual network, SN=salience network, AttN=attention network, ECN=executive control network, AN=auditory network.

| Network | Regions* | GOBS (N=334) | | HCP (N=518) | | HCP (N=481) | |
|---|---|---|---|---|---|---|---|
| | | Heritability# | p-value | Heritability# | p-value | Heritability# | p-value |
| DMN | $r_1$-$r_2$ | **0.34 (0.16)** | 0.014 | **0.27 (0.09)** | $1.0\times10^{-7}$ | **0.28 (0.09)** | $1.0\times10^{-7}$ |
| | $r_2$-$r_3$ | 0 | 0.500 | **0.14 (0.09)** | 0.008 | **0.13 (0.09)** | 0.014 |
| | $r_3$-$r_1$ | 0.09 (0.15) | 0.276 | **0.15 (0.11)** | 0.025 | 0.12 (0.11) | 0.059 |
| | $r_2$-$r_1$ | 0.09 (0.13) | 0.244 | **0.27 (0.09)** | $4.6\times10^{-8}$ | **0.25 (0.09)** | $2.0\times10^{-7}$ |
| | $r_3$-$r_2$ | 0 | 0.500 | **0.23 (0.12)** | 0.002 | **0.21 (0.12)** | $5.7\times10^{-3}$ |
| | $r_1$-$r_3$ | **0.35 (0.17)** | 0.014 | 0.09 (0.1) | 0.120 | 0.08 (0.09) | 0.094 |
| FPN | $f_1$-$p_1$ | 0.14 (0.14) | 0.149 | **0.16 (0.11)** | 0.019 | **0.16 (0.10)** | 0.018 |
| | $p_1$-$f_1$ | 0.13 (0.14) | 0.169 | **0.16 (0.11)** | 0.018 | **0.13 (0.11)** | 0.044 |
| | $f_2$-$p_2$ | **0.31 (0.15)** | 0.016 | **0.19 (0.14)** | 0.034 | **0.26 (0.14)** | 0.021 |
| | $p_2$-$f_2$ | **0.29 (0.15)** | 0.025 | **0.27 (0.14)** | 0.042 | **0.36 (0.13)** | 0.009 |
| SMN | $m_1$-$m_2$ | 0.09 (0.14) | 0.255 | **0.29 (0.15)** | 0.017 | **0.35 (0.14)** | 0.006 |
| | $m_2$-$m_3$ | 0 | 0.500 | 0.14 (0.12) | 0.113 | 0.14 (0.13) | 0.135 |
| | $m_3$-$m_1$ | **0.32 (0.20)** | 0.041 | **0.27 (0.14)** | 0.009 | **0.32 (0.14)** | 0.007 |
| | $m_2$-$m_1$ | 0.06 (0.12) | 0.302 | 0 | 0.5 | 0 | 0.5 |
| | $m_3$-$m_2$ | 0 | 0.500 | 0.15 (0.13) | 0.108 | **0.24 (0.14)** | 0.044 |
| | m$_1$-m$_3$ | **0.32 (0.20)** | 0.045 | **0.25 (0.13)** | 0.008 | **0.32 (0.14)** | 0.005 |
| VN | v$_1$-v$_2$ | 0.210 (0.15) | 0.062 | **0.14 (0.09)** | 0.021 | **0.15 (0.09)** | 0.019 |
| | v$_2$-v$_3$ | **0.36 (0.14)** | 0.004 | **0.17 (0.11)** | 0.029 | **0.20 (0.11)** | 0.017 |
| | v$_3$-v$_1$ | 0.12 (0.14) | 0.168 | 0.03 (0.04) | 0.191 | 0.05 (0.06) | 0.145 |
| | v$_2$-v$_1$ | **0.32 (0.15)** | 0.009 | **0.13 (0.09)** | 0.042 | **0.18 (0.11)** | 0.017 |
| | v$_3$-v$_2$ | 0.13 (0.14) | 0.161 | **0.15 (0.09)** | 0.040 | **0.19 (0.11)** | 0.017 |
| | v$_1$-v$_3$ | 0.17 (0.14) | 0.100 | 0.06 (0.05) | 0.060 | **0.09 (0.06)** | 0.030 |
| SN | r$_1$-r$_2$ | 0.20 (0.13) | 0.062 | 0.07 (0.09) | 0.121 | 0.06 (0.08) | 0.166 |
| | r$_2$-r$_3$ | **0.24 (0.12)** | 0.019 | **0.25 (0.11)** | 0.002 | **0.28 (0.13)** | 0.002 |
| | r$_3$-r$_1$ | 0 | 0.5 | **0.13 (0.08)** | 0.005 | **0.12 (0.08)** | 0.013 |
| | r$_2$-r$_1$ | 0.16 (0.12) | 0.084 | **0.20 (0.11)** | 0.002 | **0.17 (0.11)** | 0.008 |
| | r$_3$-r$_2$ | **0.18 (0.12)** | 0.049 | **0.31 (0.12)** | $3.8\times10^{-4}$ | **0.32 (0.12)** | $4.0\times10^{-4}$ |
| | r$_1$-r$_3$ | 0 | 0.5 | 0.05 (0.06) | 0.142 | 0.04 (0.06) | 0.196 |
| AttN | f$_1$-p$_1$ | **0.20 (0.12)** | 0.031 | 0.10 (0.15) | 0.288 | 0.08 (0.19) | 0.384 |
| | p$_1$-f$_1$ | **0.21 (0.12)** | 0.024 | 0.08 (0.11) | 0.213 | 0.05 (0.10) | 0.274 |
| | f$_2$-p$_2$ | **0.32 (0.12)** | 0.001 | **0.27 (0.14)** | 0.018 | **0.32 (0.14)** | 0.011 |
| | p$_2$-f$_2$ | **0.31 (0.12)** | 0.002 | **0.32 (0.14)** | 0.005 | **0.35 (0.14)** | 0.004 |
| ECN | r$_1$-r$_2$ | 0.17 (0.14) | 0.088 | **0.17 (0.11)** | 0.023 | **0.18 (0.12)** | 0.015 |
| | r$_2$-r$_1$ | **0.23 (0.14)** | 0.034 | **0.23 (0.11)** | 0.003 | **0.22 (0.11)** | 0.002 |
| AN | a$_1$-a$_2$ | 0.12 (0.16) | 0.209 | 0.05 (0.09) | 0.275 | 0.03 (0.07) | 0.336 |
| | a$_2$-a$_1$ | 0.05 (0.14) | 0.356 | 0.05 (0.08) | 0.260 | 0.04 (0.08) | 0.303 |

Table 2. Heritability estimates for measures derived from DMN (GOBS). * Regions are based off of **Figure 3.** Bolded connections are significant after multiple comparisons correction with FDR at q=5%. #Estimated heritability, h$^2$ (SE).

| Network | | Seed-based approach | | Dual regression approach | |
|---|---|---|---|---|---|
| | Regions* | Heritability# | p-value | Heritability# | p-value |
| DMN | r$_1$-r$_2$ | **0.34 (0.16)** | 0.014 | **0.30 (0.17)** | 0.027 |
| | r$_2$-r$_3$ | 0 | 0.500 | 0 | 0.5 |
| | r$_3$-r$_1$ | 0.09 (0.15) | 0.276 | 0.09 (0.11) | 0.279 |
| | r$_2$-r$_1$ | 0.09 (0.13) | 0.244 | **0.26 (0.16)** | 0.035 |
| | r$_3$-r$_2$ | 0 | 0.500 | 0 | 0.5 |
| | r$_1$-r$_3$ | **0.35 (0.17)** | 0.014 | **0.25 (0.17)** | 0.038 |

## 4. Discussion

We applied the ENIGMA rsfMRI pipeline to two datasets (GOBS and HCP), collected ten years apart, to demonstrate that we could consistently detect genetic influences on resting state connectivity. Building on prior work in individual cohorts, this experiment provides direct evidence that connectivity within the DMN and other intrinsic brain networks is influenced by genetic factors. We found between 20-40% of the intersubject variance in functional connectivity within functional networks was under genetic control. Our findings replicate previously reported heritability measurements in the GOBS cohort and extend this research by conducting harmonized analyses in the HCP subjects. The pattern of heritability was similar between two cohorts collected using very different imaging protocols and sample designs. Together, these findings strongly suggest that resting state connectivity is under a moderate genetic control and this heritability can be detected in, in terms of image acquisition, both legacy and state-of-the art samples. Establishing the consistency of the heritability of resting state functional connectivity provides critical information necessary before these measures can be appropriately used in genetic studies designed to identify or functionally characterize genes influencing measures of brain function. Showing reproducible and significant heritability is necessary before indices of default-mode functional connectivity can be considered as an intermediate phenotype or endophenotype for in-depth genetic analyses.

The ENIGMA rsfMRI pipeline is built using the NIH-supported software - AFNI - that is freely available to both non-commercial and commercial users. The use of free-license software opens ENIGMA collaboration to commercial entities such as pharmacological companies. It is a unimodal analysis workflow designed for consistent retrospective analyses of state-of-the-art and legacy data. The pipeline incorporates stringent quality assurance (QA) and QC steps. It incorporates traditional QA measurements to detect and censor motion and other types artifacts that are detectable visually. It also uses novel analysis of the heterogeneity of the thermal noise within imaging volume to enable identification of more subtle artifacts such as time-and-space related variability in the coil sensitivity profiles.

**Acknowledgments**

Support was received from NIH grants U54EB020403, U01MH108148, 2R01EB015611, R01MH112180, R01DA027680, R01MH085646.